\documentclass[vecphys]{svmult}

\usepackage{makeidx}         
\usepackage{graphicx}        
\usepackage{multicol}        
\usepackage{cite}            
\usepackage[bottom]{footmisc}

\makeindex             

\begin{document}

\title{Exciton-polariton condensates in zero-, one-, and two-dimensional lattices}
\author{Na Young Kim${^1}$, Yoshihisa Yamamoto${^{1,2}}$, Shoko Utsunomiya${^2}$, Kenichiro Kusudo${^2}$, Sven H\"ofling${^3}$, \and Alfred Forchel${^3}$}
\institute{${^1}$E. L. Ginzton Laboratory, Stanford University, Stanford California, 94305 USA
\texttt{nayoung@nayoung@stanford.edu}
${^2}$National Institute of Informatics, Hitotsubashi, Chiyoda-ku, Tokyo 101-8430, Japan ${^3}$ Technische Physik and Wilhelm-Conrad-R\"ontgen-Research Center for Complex Material Systems, Universit\"at W\"urzburg, 97074, W\"urzburg, Am Hubland, Germany}

\maketitle

\begin{abstract}

Microcavity exciton-polaritons are quantum quasi-particles arising from the strong light-matter coupling. They have exhibited rich quantum dynamics rooted from bosonic nature and inherent non-equilibrium condition. Dynamical condensation in microcavity exciton-polaritons has been observed at much elevated temperatures in comparison to ultrocold atom condensates. Recently, we have investigated the behavior of exciton-polariton condensates in artificial trap and lattice geometries in zero-dimension, one-dimension (1D) and two-dimension (2D). Coherent $\pi$-state with \textit{p}-wave order in a 1D condensate array and \textit{d}-orbital state in a 2D square lattice are observed. We anticipate that the preparation of high-orbital condensates can be further extended to probe dynamical quantum phase transition in a controlled manner as quantum emulation applications.
\end{abstract}

\noindent

\section{Overview}

All optical phenomena, whether visible or invisible, are ultimately resulting from an underlying mechanism: light-matter interaction. As a form of electromagnetic waves, the static and dynamical behavior of light and its interaction are well described within the classical electromagnetism framework, which was established more than 150 years ago~\cite{Jackson}. Combined with quantum pictures, this fundamental physical knowledge conceptualized spontaneous and stimulated emission of radiations early 20th century, and the quest of engineering the light-matter interaction has been a driving force to invent novel and influential photonic devices.  An optical cavity or optical resonator is an essential structure to confine the light and modify its interaction. The simplest form of the cavity is a pair of mirrors, wherein resonant standing waves are formed from multiple reflections off the mirror surfaces. By locating the gain medium inside the cavity with respect to the light standing waves, light-matter coupling is readily manipulated, consequently enhancing or suppressing spontaneous and stimulated emission~\cite{Haroche1989}. Recently, compact micron-size cavities, so-called microcavities, have been developed where the overlap between the light and matter is greatly increased. Numerous crafted microcavities  of high quality factor \textit{Q} are designed to incorporate with single or ensembles of solid-state light emitters, leading to influential optical applications and fundamental research activities in cavity quantum electrodynamics~\cite{Vahala2003}.

Among the ingenious designs of cavities, a planar Fabry-Perot resonator which consists of two mirrors enjoys simplicity and versatility to combine with different forms of matter. In particular, dielectric Fabry-Perot mirrors created by alternating two different refractive-index semiconductors can make monolithic structures combining with semiconductor gain media, for example, embedded quantum dots and quantum wells (QWs). When the gain media resides in a designated position, antinodes of the confined electromagnetic field  distribution inside the cavity, the light and matter exchange energy reversibly, reaching the strong coupling regime. This chapter focuses on strongly coupled microcavity photons with QW excitons in III-V GaAs based semiconductors. In particular, it describes the recent research activities to investigate emergent quantum phases appearing in  microcavity exciton-polariton condensates trapped in artificial lattice potentials.

\subsection{Microcavity Exciton-Polaritons and Condensation}

As cavity photons and QW excitons are strongly interacting through multiple reversible energy exchanges, new quantum quasi-particles, exciton-polaritons, emerge~\cite{Weisbuch1992}. Mathematically, exciton-polaritons are eigenmodes of the coupled cavity photon-QW exciton Hamiltonian $\hat{H}$. The Hamiltonian is written as a second quantization format in terms of the cavity photon operator $\hat{a}_k$ with energy $\hbar \omega_{ph}$ and the QW exciton operator $\hat{C}_k$ with $\hbar \omega_{exc}$ and their inbetween interaction coupling constant $g_k$,

\begin{equation}
\hat{H} =\hbar \sum_k [\omega_{ph}\hat{a}^\dagger_k\hat{a}_k + \omega_{exc}\hat{C}^\dagger_k\hat{C}_k - ig_k(\hat{a_k}^\dagger\hat{C}_k-\hat{a}_k\hat{C}^\dagger_k)].
\end{equation}

This Hamiltonian is diagonalized with an exciton-polariton operator at a momentum $k$, $\hat{P}_k = u_k\hat{C}_k + v_k\hat{a}_k$, a linear superposition of cavity photon and QW exciton operators. The resulting Hamiltonian $\hat{H}_T$ is simplified to

\begin{equation}
\hat{H}_T = \sum_k \hbar \Omega_k \hat{P}_k^\dagger\hat{P_k},
\end{equation}
\noindent and exciton-polariton frequency relations are given by

\begin{equation}
\Omega_k = \frac{1}{2}(\omega_{exc}+\omega_{ph}) \pm \frac{1}{2}\sqrt{(2g_k)^2+(\omega_{exc}-\omega_{ph})^2}.
\end{equation}
\noindent This strong coupling manifests unequivocally as a pronounced energy doublet (upper polariton (UP) and lower polariton (LP)  branches), whose energy separation indicates the coupling interaction strength ($2g$) denoted as vacuum Rabi splitting along the same spirit of the atom-cavity systems~\cite{Kavokin, Yamamoto, Deng2010}. 

The dual nature of microcavity exciton-polaritons provides advantages to explore fundamental quantum Bose nature~\cite{Kavokin, Yamamoto} and to engineer potential photonic and optoelectronic devices~\cite{Deveaud2008, Liew2011}. Elaborately, the partial photonic nature reduces the effective mass of this composite particle down to about 10$^{-4} \sim 10^{-5}$ of the electron mass and about 10$^{-8}$ of the hydrogen atom mass. The extremely light effect mass of particles makes us to easily execute experiments at high operating temperatures. There are abundant photon flux leaked from the cavity structure owing to the finite lifetime of exciton-polaritons. These leaked photons carry out the energy-momentum distribution of exciton-polaritons under the energy and momentum conversation. Hence, capturing those leaked photons enables us to access polariton dynamics. On the other hand, electrons and holes, fermionic constituents of QW excitons, are Coulombically interacting and their repulsive interaction plays a significant role in scattering processes. These non-zero interactions among particles enrich the phase diagram of microcavity exciton-polaritons. 

Being composite bosons of photons and excitons, exciton-polaritons in the low density limit and at low enough temperatures are predicted to reveal novel quantum Bose nature such as Bose-Einstein condensation (BEC) utilizing bosonic final state stimulation and stimulated scattering processes~\cite{Imamoglu1996}. During last two decades since the discovery~\cite{Weisbuch1992}, the increased interest of microcavity exciton-polariton condensates has led to tremendous advancement both in theory and experiments, exploring unique BEC nature in microcavity exciton-polaritons. Several groups have reported strong evidences of exciton-polariton BEC in terms of macroscopic occupation in a ground state, thermal equilibrium to lattices, spontaneous long-range spatial and increased temporal coherence properties ~\cite{Deng2002, Deng2003, Kasprzak2006, Deng2006, Deng2007, Balili2007, Lai2007}.

Exciton-polariton condensates have exhibited distinct features from atomic counterparts in several aspects. First, the reduced effective mass results in elevated BEC phase transition temperatures, which are inversely proportional to the mass. GaAs and CdTe systems undergo the phase transition around 4 - 10 K~\cite{Deng2002,Deng2003, Kasprzak2006, Balili2007}, and large bandgap materials like GaN and organic systems show BEC at room temperatures~\cite{Christopoulos2007, Christmann2008, Cohen2010}, 10$^8 \sim 10^9$ times higher than the transition temperatures (tens of nK) of atomic BECs. Second, the macroscopic condensate population at the ground state can be accumulated by constant injection of particles in order to compensate leakages due to short quasi-particle lifetime. This open-dissipative nature is responsible for unique dynamics of exciton-polariton condensates.

Rigorously, owing to the fact that microcavity exciton-polaritons reside in two-dimension (2D), in the thermodynamical limit, infinite 2D systems cannot exhibit BEC at non-zero temperatures due to phase fluctuations in principle according to Hohenberg theorem~\cite{Mermin1966, Hohenberg1967}. However, in finite 2D systems, long-range off-diagonal coherence can still be preserved through the Berezinskii-Kosterlitz-Thouless (BKT) transition~\cite{Berezinskii1971,Kosterlitz1973, Kosterlitz1974}, where macroscopic quantum coherence would be stabilized by forming vortex-antivortex pairs. This fundamental inquiry has motivated to study trapped exciton-polariton condensates in spatial in-plane potentials since a well-defined trapping potential profile would guarantee single-mode BEC at discretized energy states. The following subsection summarizes attempts to create various types of trapping potentials in exciton-polariton systems.

\subsection{Types of In-plane Trapping Potential}

\begin{figure}[t]
\centering
\includegraphics*[width=.7\textwidth]{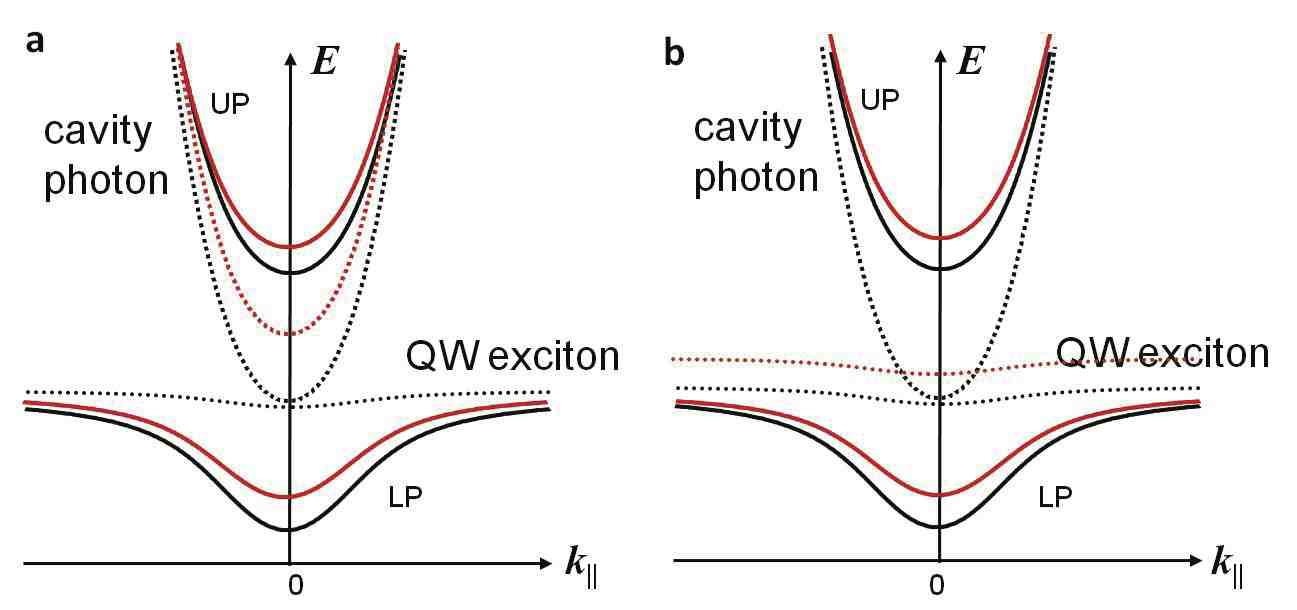}
\caption[]{Modified upper and lower polariton (UP, LP) energy dispersion relations (red straight lines) by shifting either cavity photon (a) or quantum well (QW) exciton modes (b) indicated by red dotted lines. Black straight lines draw the original UP and LP energy dispersions arising from the cavity photons and QW excitons drawn in the black dotted lines}
\label{fig:1}       
\end{figure}

Disorder, impurities and imperfections in semiconductor materials form locally isolated traps. While these natural traps are difficult to be controlled, several schemes to engineer local in-plane spatial traps have been attempted benefitting from the dual light-matter nature of exciton-polaritons. Figure ~\ref{fig:1} depicts how to modify the polariton branches correspondingly by influencing either the exciton mode energy or the cavity photon mode energy. The details of natural traps and engineered trapping potential are given in this subsection.

\paragraph{Natural Trap}
Natural spatial traps in semiconductors unavoidably exist due to semiconductor monolayer thickness fluctuations, defects and disorders. These local traps appear more often in  II-VI semiconductors than GaAs materials since GaAs materials are relatively cleaner and purer. Although the controllability of these traps is absent, rather strong confinement potential (a few meV) profile leads to many interesting phenomena: to name a few, quantized vortex pinned at disorder~\cite{Lagoudakis2008} and half-quantum vortex~\cite{Lagoudakis2009}. Sanvitto and colleagues have identified exciton-polariton condensates populated in discrete modes only via time- and energy-resolved images~\cite{Sanvitto2009}. These quantized states are from a $\sim$6 $\mu$m trap of $\sim$2 meV strength associated with one monolayer cavity thickness fluctuation.

\begin{figure}[t]
\centering
\includegraphics*[width=.7\textwidth]{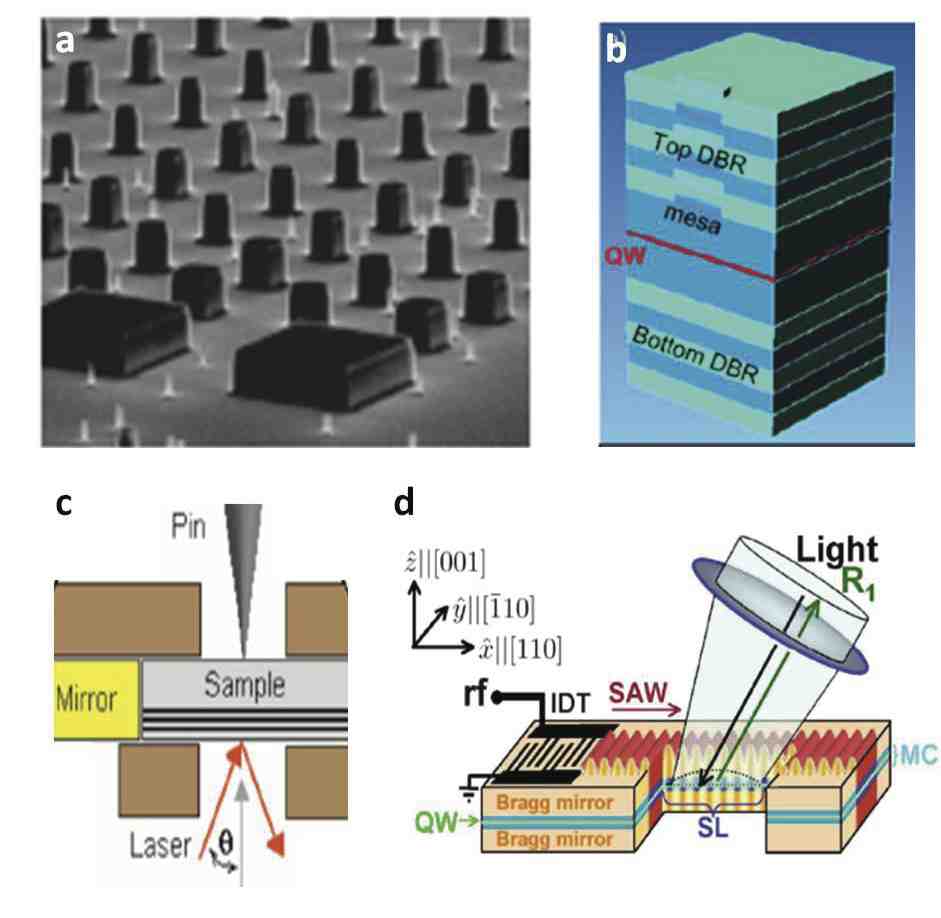}
\caption[]{a, Image of pillar arrays adapted from Ref. ~\cite{Bajoni2008}. b, Partial etching and overgrowth technique developed in Professor B. Deveaud-Pl\'{e}dran group in Ref. ~\cite{Nardin2010}. c, A schematic of a mechanical stress setup from Ref.~\cite{Balili2006}. d, Illustration of an acoustic lattice in Ref. ~\cite{deLima2006}. Figures a,b, and d have acquired copyright permissions from America Physical Society journals. Copyright 2008, 2010, 2006, American Physical Society. Figure c is reprinted with permission from Balili et al., Applied Physics Letters \textbf{88}, 031110. Copyright 2006, American Institute of Physics}
\label{fig:2}       
\end{figure}

\paragraph{Gain-Induced Trap}

Another effective confinement potential is formed due to the finite excitation laser spot size~\cite{Roumpos2010}. This implicit mechanism arises from the interlay of the finite lifetime and the spatial gain modulation bounded by the laser spot. Reservoir polaritons are rather short-lived to remain inside the finite laser spot region, building up the density and reaching the final state stimulation above quantum degeneracy threshold values. In this implicit effective trap, energy states are clearly quantized both in real- and momentum- spaces, and the Heisenberg-limited real- and momentum space distribution is observed. The experimental observations are consistent with the results of the infinite barrier circular trap model.  Notwithstanding the successful description of the zero-dimensional confinement features, this method suffers from the limitation of the laser spot size control and cannot be readily scalable.

\paragraph{Etched Structures}

The aforementioned traps exist associated with material purity and the excitation scheme, lacking of controllability and scalability. To overcome the limitation, several schemes have been contrived. As a simplistic method, micro-sized pillars or photonic dots were fabricated by an etching method for a transverse spatial confinement~\cite{Bloch1998, Gutbrod1998, Obert2004}. Modifying the photonic component spatially, the large refractive index discontinuity is induced at the pillar sidewalls. Figure ~\ref{fig:2}\textbf{a} images the array of pillars with varying sizes, which confirms that scalable polariton systems can be prepared in this etching technique. The micropillar structures clearly quantize photonic modes, consequently polariton modes as well. However, these earlier attempts failed to preserve the strong coupling regime at high excitation powers. Recently, improved etching processes enable to show confined condensates in 2-20 $\mu$m-sized micropillar cavity structures~\cite{Bajoni2008, Ferrier2011}. Not only is the quantized energy states observed as a zero-dimensional condensate confinement but also condensed polaritons reside near the pillar edges because of the strong repulsive interactions. In addition, 1D wire cavities~\cite{Wertz2010} are patterned, where condensed polaritons are extended away from the excitation pump spot, manifesting strong correlation effects. Towards the 2D extension, two-by-two photonic molecules are prepared, where condensed states form bonding and antibonding states analogous to chemical molecule boding configurations~\cite{Galbiati2011}.

Having a potential issue of the sidewall quality as a source of exciton loss and photon field leakage in micropillar structures, a slightly modified method has been implemented~\cite{ElDaif2006, Kaitouni2006, Nardin2010}. Instead of etching the whole structure, only the cavity space layer is partially etched by $\sim$6 nm. The height profile continues to be transferred to the subsequent layers after  the overgrowth of the remaining structure (Fig. ~\ref{fig:2}\textbf{b}). This physical cavity length spatial modulation produces strong confinement potential of  $\sim$9 meV.  The 3-20 $\mu$m-diameter mesa structures behave polariton quantum dots with discrete polaritonic energy spectra modified by confined photon modes. The nice thing of this alternative attempt lies in the fact that the partial etching does not touch QWs so that any spurious effect like surface recombination would be negligible. Incorporating with semiconductor processing techniques, this method would have flexibility to pattern arrays of mesa structures even in consideration of coupling control between nearest neighbor mesas.

\paragraph{Temperature and Electric Field Tuned Trap}

Whereas both the photonic disorder and the etching influence the cavity photon mode, there is another route to produce in-plane trapping potentials: shifting the QW exciton mode.  Since  the  exciton mode is more sensitive to the environment temperatures through material lattice constant variations than the photon mode,  temperature can tune the exciton energy and hence the polariton energy as well~\cite{Fisher1995}. However, since temperature affects the whole device, it is not possible to engineer local trap potentials unless a delicate approach is introduced.  Another tunable way is to apply electric fields through the quantum-confined Stark effect~\cite{Miller1985}. Basically, the non-zero electric fields reduce the overlap of electron and and hole wavefunctions in QWs. As a result, and the shifted exciton modes are mixed with cavity photons and the smaller oscillator strength is measured~\cite{Fisher1995}. Electric-field tuned traps can be scalable by lithographic patterns and can be controlled in situ by varying the applied field values; however, to our best knowledge, trapped condensation inside the electrostatic traps has not been reported yet.

\paragraph{Stress Trap}

Another way to affect the exciton mode in the microcavity-QW structure is to apply mechanical stress.  A strain harmonic potential of $\sim$ 3 meV with $\sim$ 1 N force is induced and its schematic sketch is drawn in Fig. ~\ref{fig:2}\textbf{c}~\cite{Balili2006, Balili2007, Balili2009}. The harmonic potential alters the available density of states, and it assists to reduce the required particle density necessary for condensation. Since the stress-induced potential is rather large-sized $30-40 \mu$m due to the rounded-tip pin size (radius 50 $\mu$m) and the back substrate thickness ($\sim$ 100 $\mu$m), no discrete modes have been observed yet in this manner. However, exciton-polariton condensates in this wide trap exhibit several clear BEC phenomena~\cite{Balili2007, Balili2009, Nelsen2009}, and the trapped condensates would be a good candidate to explore BEC-BKT crossover in future. One concerns that it is difficult to imagine how to scale up the stress traps to arrays of them.

\paragraph{Acoustic Trap}

All above trapping potentials are time-independent and static, but dynamical traps would provide an interesting test-bed to explore condensate properties. A theoretical proposal envisages the acoustic lattices produced by strong
exciton-phonon interaction~\cite{Ivanov2001}. Rayleigh surface acoustic waves (SAWs) launched in piezoelectric GaAs-based microcavity structures produce dynamical phonon superlattices~\cite{deLima2006} (Fig. ~\ref{fig:2}\textbf{d}). The primary mechanisms of the acoustic lattices are the type I band-gap deformation and the in-phase cavity resonance energy modulation~\cite{deLima2006, Cerda2010}. The lattice constants of the phononic superlattices  are determined by the SAW wavelength. The 5.6 $\mu$m and 8 $\mu$m SAWs are driven by a conventional GHz microwave technology~\cite{deLima2006, Cerda2010}. Microcavity exciton-polariton condensates are fragmented into arrays of 1D phonon traps located at the SAW minima, exhibiting the 1D band structures by a $\sim$ 160 $\mu$eV acoustic lattice~\cite{Cerda2010}. Although dynamical acoustic lattices can be one- and two-dimensional and can be in-situ controllable, they face  challenges: how to handle microwave heating for strongest traps and how to configure various lattice geometries beyond the simple 1D array or 2D square geometry.

\paragraph{Metal-film Trap}

\begin{figure}[t]
\centering
\includegraphics*[width=.7\textwidth]{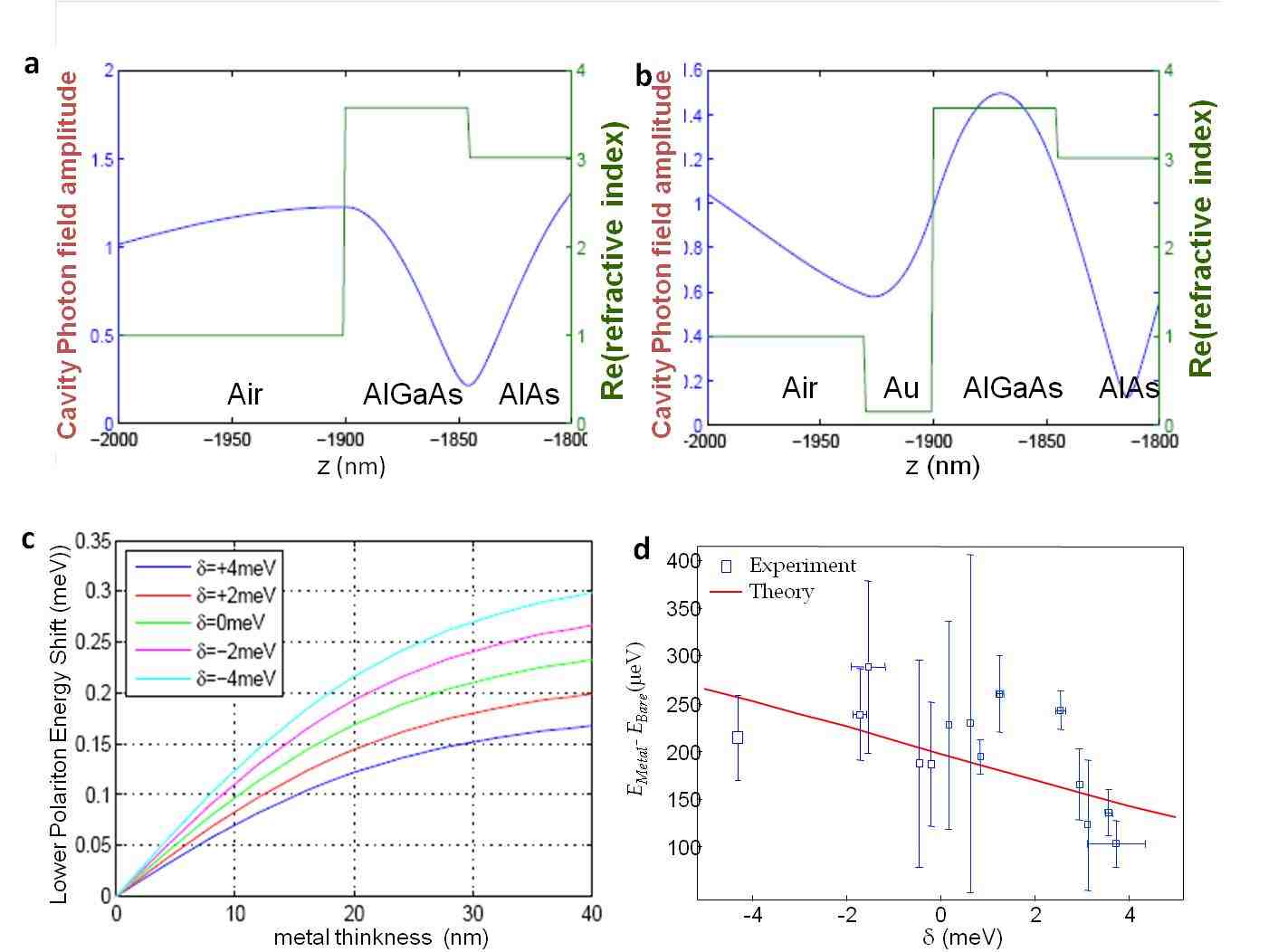}
\caption[]{The real part of refractive indices of materials (green) and the electromagnetic field amplitudes (blue) are plotted along the growth direction in $z$ by the transfer matrix calculation for a standing wave at a resonance from the bare surface (\textbf{a}) and from the 30 nm Au metal layer (\textbf{b}). c, Lower polariton (LP) energy shifts are shown as a function of the thin metal film thickness in the 16-20 top-bottom distributed Bragg reflector structure at varying detuning values $\delta$. d, The relation of the LP energy shift and the detuning values are measured with the a 24 nm-6nm Au/Ti film deposited GaAs microcavity structure}
\label{fig:3}       
\end{figure}

Finally, the concept of the metal-film traps is introduced as a new and simple way to create static in-plane potentials. A thin-metal film deposited on grown GaAs microcavity-QW wafers affects only the photonic mode ~\cite{Lai2007, Kim2008, Kim2011}.  Figure~\ref{fig:3}\textbf{a} and \textbf{b} compare the transfer matrix calculation result of the electromagnetic fields at the metal-semiconductor interface with that of the air-semiconductor interface ~\cite{Yeh}. The photon fields are extended rather smoothly at the air-semiconductor interfaces (Fig.~\ref{fig:3}\textbf{a}), but the confined electromagnetic components have to be zero at the metal-semiconductor interface (Fig.~\ref{fig:3}\textbf{b}). Therefore, the stiff boundary condition on the photonic mode due to the metal film gives rise to a locally squeezed cavity length, shifting the polariton energy upward in comparison to the polariton energy value under the air-semiconductor interfaces. With the 30 nm Au film, the cavity photon energy is blue shifted by $\sim$ 400 $\mu$eV, and the LP energy shift becomes half near the zero detuning area, where the photonic concentration is 50 \%. In the calculation, we assume the wafer of the 16 top and 20 bottom distributed Bragg reflectors (DBRs).

Figure~\ref{fig:3}\textbf{c} explicitly draws the LP energy shift versus the metal film thickness at varying detuning parameters $\delta = E_{cav} (k_{//}=0) - E_{exc} (k_{//} = 0)$, where $E_{cav(exc)}(k_{//}=0)$ denote the energy of cavity (exciton) at the zero in-plane momentum $k_{//}$. The amount of LP energy shift becomes slowly saturated in cases with more than 30nm thick metal layers. At a given metal thickness, the resulting LP energy shift is proportional to the photon concentration. The more negative $\delta$ is, the LP energy shift is bigger. Suppose the metal film induces the cavity photon energy shift by $\Delta E_{cav}$.  Then, the overall LP energy at $k_{//}=0$ below the metal film is written as
\begin{equation}
E_{metal, LP} = \frac{1}{2}[E_{cav} + \Delta E_{cav} + E_{exc}- \sqrt{(2g)^2+(E_{cav}-E_{exc})^2}].
\end{equation}
In terms of the detuning $\delta$, the LP energy difference between the metal and the bare surface is derived as
\begin{equation}
E_{metal, LP} - E_{bare, LP} = \frac{1}{2}[\Delta E_{cav}- \sqrt{(2g)^2+(\delta+\Delta E_{cav})^2}+\sqrt{(2g)^2+\delta^2}].
\end{equation}
We have measured the LP energy difference with the 26/4nm Au/Ti metal film on top of the 16/20 DBR structures in Fig. ~\ref{fig:3}\textbf{d}, which matches qualitatively well with the computed results (red line) using $g \sim 6.97$ meV, $E_{exc} \sim 1.59241$ eV and $\Delta E_{cav} \sim 400 \mu$eV.

Although the trapping potential strength is much weaker than the above etched pillars and natural traps, this method possesses several advantages: (1) the non-invasive in-plane potentials can be introduced on top of the fully grown wafer, which does not change the QW quality and exciton states; (2) the dimension of the in-plane potentials can be readily controlled even down to sub-microns by utilizing advanced lithographic techniques;  and (3) the flexibility to design various geometries in 1D and 2D is very attractive, especially when many-body interaction effects among exciton-polaritons are explored in lattices.


The main body of the chapter is devoted to describe the detailed phenomena of the microcavity exciton-polariton condensates confined in a single trap, and 1D and 2D lattices prepared by the aforementioned thin metal-film technique.

\section{Microcavity Exciton-Polariton Condensates in Lattices}

Before delving into the indepth description of experimental signatures in trapped microcavity exciton-polaritons employing the metal film technique, the wafer structures, the abridged fabrication procedure and measurement setup are briefly discussed in the next subsection \ref{sec:2}.

\subsection{Samples and Fabrication}
\label{sec:2}
\subsubsection{GaAs samples}
Our $\lambda$/2 cavity contains three-stack of four GaAs QWs located at the central antinodes of the microcavity optical field.  GaAs 6.8 nm-thick QWs are separated by 2.7 nm-AlAs barriers. The planar Fabry-Perot cavity is arranged by top and bottom DBRs from alternating $\lambda$/4 Al$_{0.15}$Ga$_{0.85}$As and AlAs layers. A 16-layer top and 20-layer bottom DBR structure promises the microcavity quality factor $\emph{Q} \sim 6000$ according to the transfer matrix calculation. The measured vacuum Rabi splitting energy $2g$ is around 15 meV near the zero-detuned area. A spatial inhomogeneity caused by a tapering in the layer thickness of the wafer allows us to tune the cavity resonance with respect to the exciton energy, which is relatively constant over the whole wafer.

\subsubsection{Fabrication Procedure}

On the top DBR surface of the grown microcavity wafer, we have designed circular traps, 1D metal strip gratings, and 2D square lattice. The 2D square lattice is patterned by arranging holes and we applied negative resist to the wafer followed by either electron-beam or photo-lithography depending on the feature sizes. The feature sizes vary from 1 to 10 $\mu$m. Then, a 24/6 nm Au/Ti metal film is deposited and finished with a lift-off in acetone. 

\subsubsection{Photoluminescence Setup}

Our primary measurements are to capture time-integrated photoluminescence (PL) signals of emitted LPs near 4 K in both near- and far-fields.  Typically we excite our samples with a $\sim$ 3 ps pulse ($\sim$ 0.5 meV spectral  width) at $\sim$ 60 degree ($k_{//} \approx 7 \times 10^4$ cm$^{-1}$ in air) by a mode-locked Ti:sapphaire laser at a 76 MHz repetition rate.  The laser excitation wavelength is tuned to be near the exciton resonance energy near 768 nm.  The excitation spot is focused on the sample surface, whose size is around typically 30 $\times$ 60 or 50 $\times$ 100 $\mu$m$^2$. The elliptical shape of the excitation spot is from the large angle side pumping scheme.  The LP emission signals are collected in the normal direction by a high numerical aperture (NA = 0.55) objective lens. We built a standard micro-PL setup to access near-field (coordinate space) or far-field (momentum space) imaging and spectroscopy with a repositionable lens behind the objective lens. For spectroscopic information, the collected signals are fed into a 0.75 m spectrometer, dispersed by gratings and recorded on a liquid nitrogen cooled CCD camera.

\begin{figure}[t]
\centering
\includegraphics*{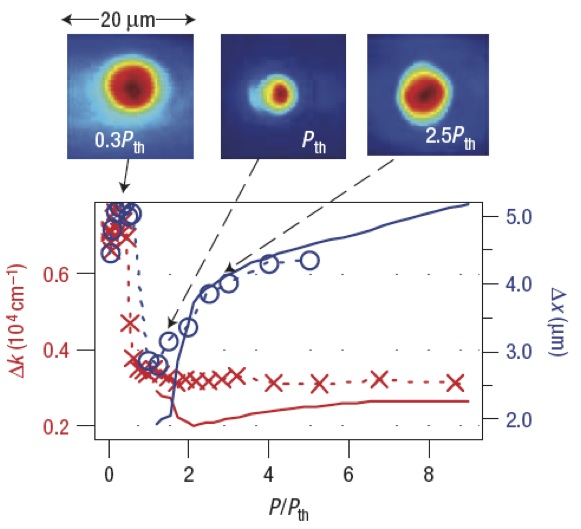}
\caption[]{Pump power-dependent near-field lower-polariton population distributions in a 8 $\mu$m-diameter single trap. The bottom panel summarizes the momentum and position standard deviations as a function of the normalized pump powers. Permission is acquired from Nature Physics}
\label{fig:4}       
\end{figure}

\subsection{Exciton-polartion condensates in a single trap}

The isolated circular traps are displayed, whose trap diameters are chosen between 5 $\mu$m and 100 $\mu$m and whose LP trapping potential strength is approximately 200 $\mu$eV.  Although the potential strength is compatible with the size of polariton kinetic energy ($\sim$ 100 - 300 $\mu$eV) in relatively small diameter-traps (5 - 10 $\mu$m), exciton-polariton condensates remain in a single transverse mode  over higher-order transverse modes shown in Fig. \ref{fig:4} ~\cite{Utsunomiya2008}.

To characterize trapped condensates, the standard deviations in momentum ($\Delta k$) and position ($\Delta x$) are recorded as a function of the excited laser pump power in the main panel of Fig. \ref{fig:4}. Both the momentum and the position fluctuations plummet abruptly at the quantum degeneracy threshold values ($P \approx P_{th}$). The measured uncertainty product $\Delta k \Delta x$ has a minimum value, 0.98 just above the threshold values. It is twice of the Heisenberg uncertainty product value ($\Delta k \Delta x \sim 0.5$) given by a minimum uncertainty wavefunction. Where $\Delta k$ remains constant above $P_{th}$, the position uncertainty $\Delta x$ significantly increases, hence the monotonic increase in $\Delta k \Delta x$. The increased condensate size directly manifests the repulsively interacting coherent LP condensates. The more polaritons are created by higher pump rates, the interaction energy among LPs increases, consequently LPs extends outward from the repelling one another. The experimental data show a consistent agreement with theoretical analysis using the Gross-Pitaevskii equation.
\begin{figure}[t]
\centering
\includegraphics*[width=.7\textwidth]{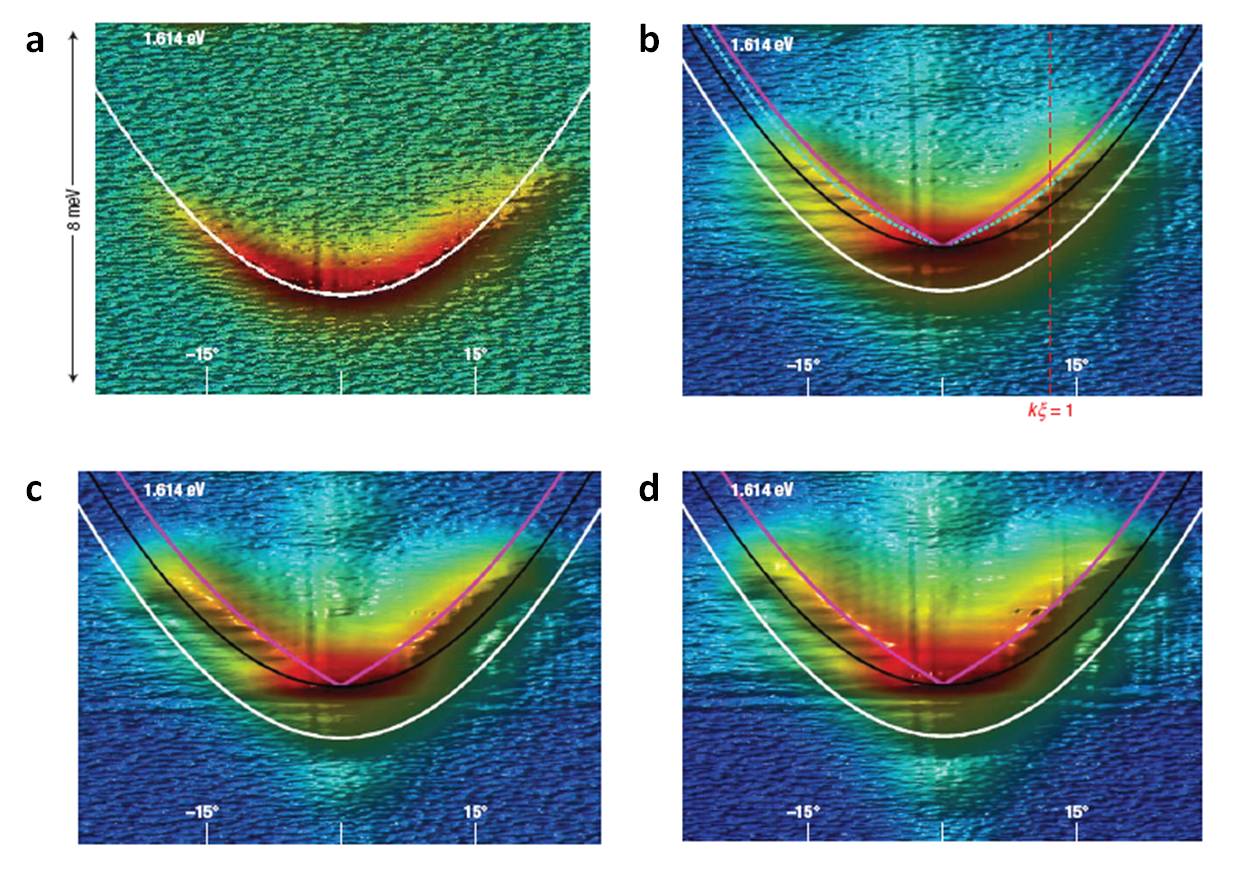}
\caption[]{Pump power-dependent energy dispersion and excitation spectra of trapped lower-polariton condensates at $P/P_{th}$ = 0.05, 1.2, 4, and 6 from \textbf{a} to \textbf{d}.  The data were taken with a 8 $\mu$m-diameter single trap, and the energy axis in y spans 8 centered at 1.61 V. The white line indicates the free-polariton parabolic energy dispersion relations, the black line is the shifted parabolic curve at the condensed state, and  the pink line draws the Bogoliubov excitation spectra computed using the homogeneous model. Permission of this figure is from Ref. ~\cite{Utsunomiya2008} is acquired from Nature Physics}
\label{fig:5}       
\end{figure}

In addition, it is natural to search the influence of the inter-particle interaction on the condensates. The excitation spectra of condensates indeed reflect the signatures of superfluidity in terms of the population fluctuations and phase stabilization of condensates~\cite{Bogoliubov1947, Pitaevskii}. Quantum and thermal depletions from the condensates excite the phonon-like spectrum. Clearly deviating from the free-particle quadratic dispersion curve, the linear Bogoliubov excitation spectrum of trapped condensates above the threshold pump powers is observed within the low-momentum regime $|k\xi|<1$. $\xi$ is the healing length given by $\xi=\hbar/\sqrt{2mU(n)}$ in terms of the effective mass $m$ and the LP density ($n$) dependent interaction strength $U(n)$. It is one of superfluidity signatures shown in Fig. ~\ref{fig:5}. The Bogoliubov energy relation $E_B$ is written as $E_B=\sqrt{E_{free}(E_{free}+U(n))}$, where $E_{free}=\hbar^2 k^2/2m$ is the free-particle kinetic energy at $k_{//}=0$. From the phonon-like linear dispersion relation, the extracted sound velocity is of the order of 10$^8$ cm/s, eight orders of magnitude faster than that of atomic condensates. Again this large sound velocity value comes from the extremely light effective mass of LPs.

\subsection{1D condensate arrays}

\begin{figure}[t]
\centering
\includegraphics*[width=.7\textwidth]{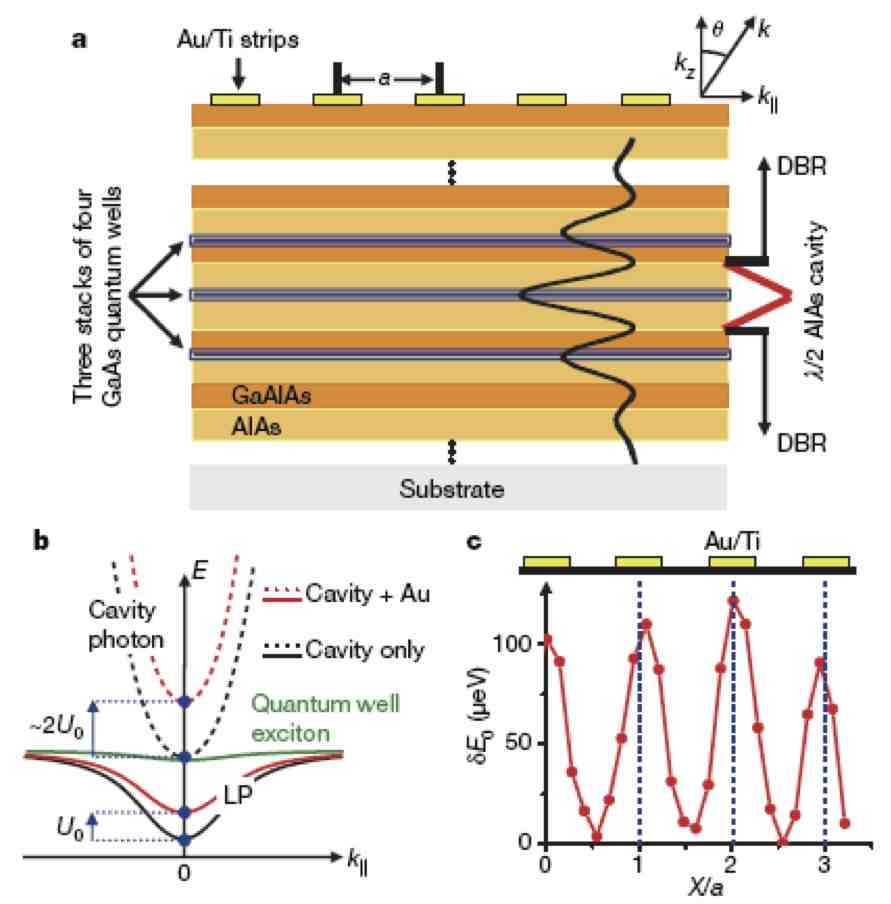}
\caption[]{a, The cross-sectional view of the one dimensional grating film on a 12 GaAs quantum-well-microcavity structure.
b, comparison of the polariton dispersions with two cases: the metal-cavity interface and the air-cavity interface. c, experimentally measured in-plane trapping potential for lower polaritons. The figure permission from Ref.~\cite{Lai2007} was given from Nature}
\label{fig:6}       
\end{figure}

In order to realize 1D condensate arrays, we have patterned a grating structure, where $a$/2-wide metal strips are equidistantly located with a $a/2$ wide gap. $a$ represents the 1D lattice period illustrated in Fig. ~\ref{fig:6}\textbf{a}, and the experimental data are taken with a device $a$ = 2.8 $\mu$m. Below the metal strip, LPs encounter higher energy barriers caused by the locally squeezed cavity lengths, shifting the LP energy by $U_0 \sim 200 \mu$eV drawn in Fig. ~\ref{fig:6}\textbf{b}. The Au/Ti metal film layer gives the spatial LP energy modulation $\delta E_0$ as  100 $\mu$eV measured by scanning the pinhole across the metal strips (Fig. ~\ref{fig:6}\textbf{c}). The reduced LP energy shift measured in Fig. ~\ref{fig:6}\textbf{c} is due to the diffraction-limited spatial resolution 1 $\mu$m, which is compatible to our device lattice constant. We have confirmed lower LP energy by 200 $\mu$eV under the bare surface independently.

\begin{figure}[t]
\centering
\includegraphics*[width=.7\textwidth]{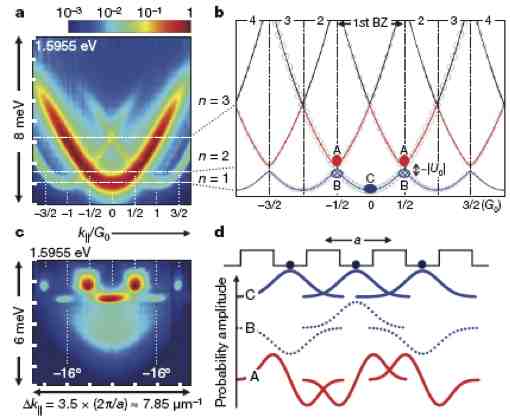}
\caption[]{a, The measured angle-resolved spectroscopy of lower polaritons (LP) trapped in the $a$ = 2.8 $\mu$m-period one-dimensional (1D) array taken at the below quantum degeneracy threshold values. $G_0$ is the reciprocal lattice vector, 2$\pi/a$. b, The single-particle 1D band structure in the extended zone scheme with the location of the first Brillouin zone. The circles represent the expected LP emission intensities. c, the energy dispersion diagram above the quantum degeneracy threshold values. d, the real-space wavefunction probability amplitudes in space. The figure is adapted from Ref.~\cite{Lai2007} with proper permission}
\label{fig:7}       
\end{figure}

The potential energy 200 $\mu$eV is of the same order of the kinetic energy $E_{free} = \hbar^2k^2/2m$ at the 1D Brillouin zone (BZ) boundaries $k_{//} = \pm \pi/a$. The kinetic energy value is $\sim$ 500 $\mu$eV with the effective mass $m$ to be $9 \times 10^{-5}$ of the electron mass. Therefore, this energy scaling comparison justifies us to treat our system as a single-particle in a weakly periodic 1D potential. We perform the 1D band structure calculation including a 200 $\mu$eV potential term. Figure ~\ref{fig:7}\textbf{b} presents the theoretical band structures against the in-plane momentum in the unit of the 1D reciprocal lattice vector amplitude $G_0 = 2\pi/a$ in the extended zone scheme. The weak potential $U_0$ lifts the doubly degenerate energy states at the first BZ edges and the two energy states are denoted as $A$ and $B$ in Fig.~\ref{fig:7}\textbf{b}. Below threshold pump powers, the dominant parabolic dispersion curves together with two relatively weaker parabolas displaced by $\pm G_0$ are detected in Fig. ~\ref{fig:7}\textbf{a}, which resemble the theoretical band structure calculation result. The spectral linewidth of the dispersion curves are bigger than the energy gap ($\sim |U_0|$) between two states $A$ and $B$ so that we were not able to directly measure the energy gap.

Above quantum degeneracy threshold values (Fig.~\ref{fig:7}\textbf{c}), the angle-resolved energy spectroscopy consists of narrow interference peaks at two different energy values as well as blurred emission from the thermal LPs below. The peaks are separated by $G_0$ (correspondingly, 16$^o$ in angles) at given energy values as a manifestation of diffraction from the 1D lattice. More important, the higher energy states occur at $\pm G_0/2$, the first BZ boundaries. The momentum distribution at this energy state is associated with the Bloch wavefunction at the $A$ state, which has anti-phased relations between nearest-neighbor sites, whereas two other Bloch states $B$ and $C$ are in phase depicted in Fig.~\ref{fig:7}\textbf{d}. The anti-phased $\pi$-state in $A$ exhibits the $p$-wave symmetry, and the in-phase $0$-states in $B$ and $C$ show the $s$-wave symmetry. Even though the state $A$ is energetically higher than the state $B$, LPs are condensed in $A$ since it is meta-stable protected by local bandstructure curvatures. The narrow emission lobes are evidences of the coherent condensates. At pump powers far above threshold values, the LP condensates indeed relax to the state $C$, which is the global ground state owing to the enhanced stimulated scattering processes among more LP particles.

Such dynamical relaxation and mode competition among multi-orbital states have been studied by the time-resolved spectroscopy.  Near the threshold the condensates are easily trapped in high-momentum, meta-stable \emph{p}-wave states initially, and the LPs in $p$-wave states decay rather faster than the ground \emph{s}-wave state. Using the reasonable lifetime parameters for these energy states, the coupled rate equations with the \emph{s}-, \emph{p}- and reservoir states would well reproduce the observed dynamical behavior of $s$- and $p$-wave condensation. In comparison to the atomic condensate, the reason to create coherent condensates at non-zero momentum values readily with the LPs can be found in the fact that LPs are short-lived so that they leak out of the cavity before relaxation into the ground states at a certain particle density range. We have also shown that the off-diagonal long-range coherence preserves the entire LP condensate sizes bounded by the laser pump spot size $\sim 50 \times 100 \mu$m$^2$. Hence, the LP system has a knob of excitation power, detuning parameters and the geometry to manipulate the condensate orbital symmetry and to search macroscopic quantum order. Next, we will describe a simple extension to the 2D square lattice using the advantage of the LP condensates.

\subsection{2D square lattice}

\begin{figure}[t]
\centering
\includegraphics*[width=.7\textwidth]{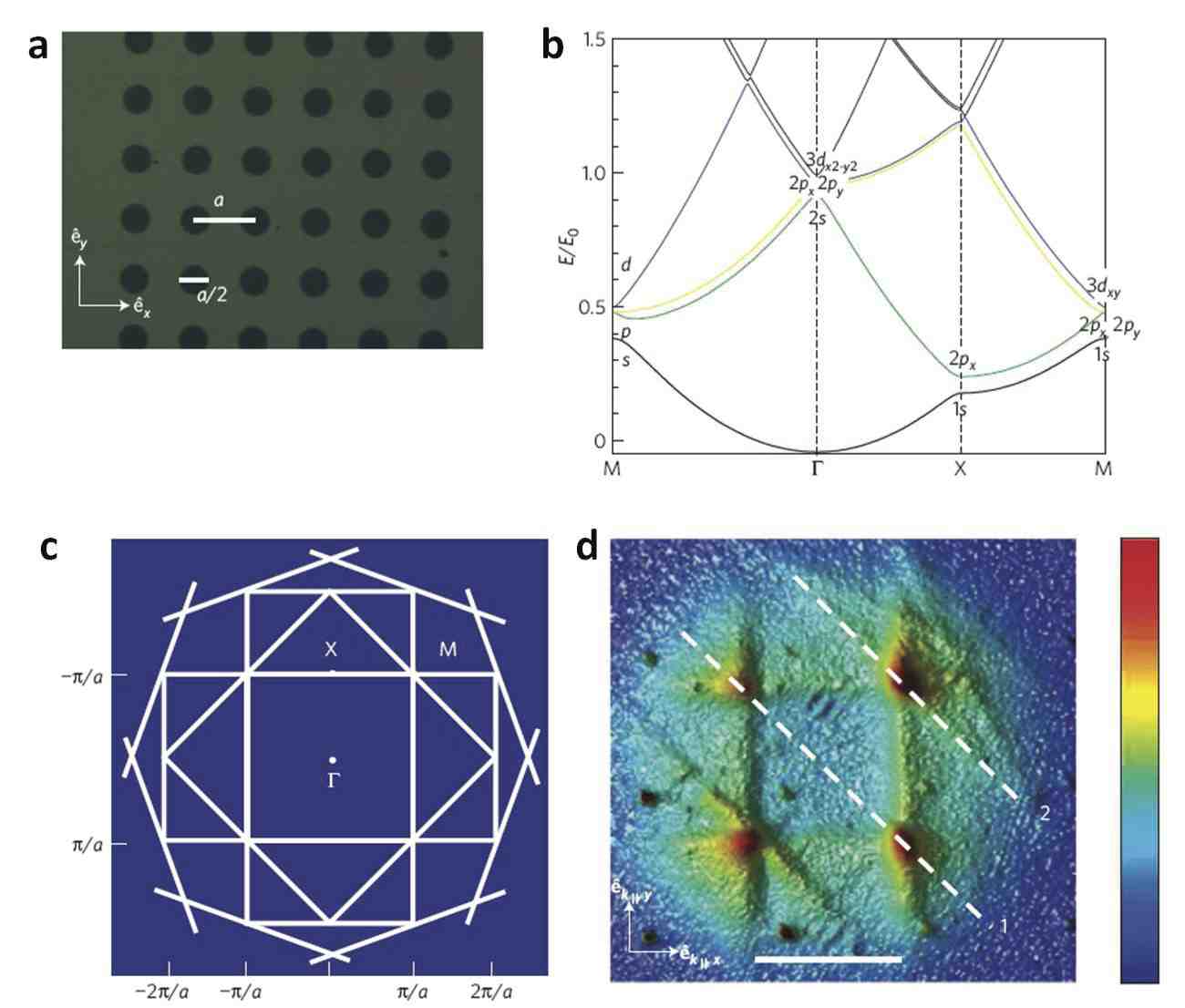}
\caption[]{a, Photograph of a representative two-dimensional (2D) square lattice device with the lattice constant $a$. The brighter area is where the metal films are deposited and the darker holes with the diameter $a/2$ are artificial lattice sites. b, The single-particle bandstructures along high symmetry points ($\Gamma, X, M$) in the presence of a weak square potential. c, The multiple Brillouin zones (BZs) of the 2D square lattices. d, Experimentally observed first and second BZs in the momentum space ($k_{//,x}, k_{//,y}$). Figure Permission is acquired from Nature Physics}
\label{fig:8}       
\end{figure}

The dimensionality of 2D is very special in that many exotic physical phenomena in strongly correlated materials are closely tied to the high orbital electronic states arranged in 2D. High temperature superconductivity is one of the prime examples, and its properties would be owing to the \emph{d}-orbital copper- and \emph{p}-orbital oxide 2D planes. Utilizing the bottleneck condensate nature, exciton-polariton condensates can be engineered to form the excited state orbital symmetry in terms of the lattice constants and the excitation pumping condition.

A square lattice is one of the simplest 2D lattice structures with orthogonal real- and reciprocal-Bravais unit vectors. The Bravais lattice forms a square first BZ with a unit length $2\pi/a$, where $a$ is the square lattice constant. The square lattice holds translational, rotational and reflection symmetries and three high symmetry points are denoted as $\Gamma, X, M$. Both $\Gamma$ and $M$ preserve the four-fold rotational symmetry, while $X$ points exhibit two-fold rotational symmetry. The fabricated 2D square lattice device is shown in Fig.~\ref{fig:8}\textbf{a}~\cite{Kim2011}.

\begin{figure}[b]
\centering
\includegraphics*[width=.7\textwidth]{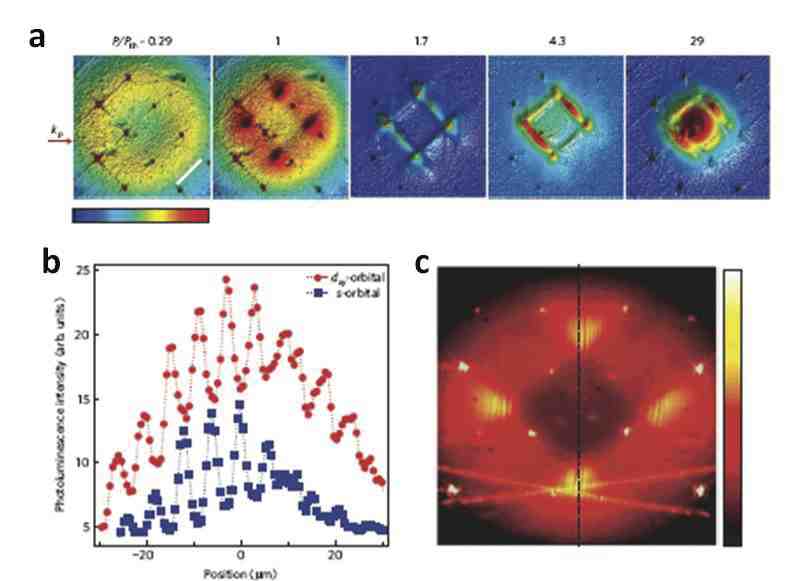}
\caption[]{a, Power-dependent momentum space lower-polariton population distributions. b, Anti-phased $d$- and in-phased $s$- wave near-space wavefunctions. c, Michelson interference images of the $d_{xy}$-orbital condensates taken at $P/P_{th} = 1$, and the folding plane is denoted as the dotted line at the center. Figure is adapted from Ref.~\cite{Kim2011} with a proper permission }
\label{fig:9}       
\end{figure}

The single-particle square lattice band structures are calculated using the plane-wave basis~\cite{Ashcroft} by solving the Schr\"odinger equation,
\begin{equation}
-\frac{\hbar^2}{2m}\nabla^2\Psi(\vec{r})+V(\vec{r})\Psi(\vec{r})= E\Psi(\vec{r}).
\end{equation}
The 2D periodic potential $V(\vec{r})$ is modeled as $V(\vec{r})=\sum_i l(\vec{r}-\vec{R_i})$ where $\vec{R_i}$ is the $i$-th site center location, and the circular potential $l(\vec{r})$ is set to be $- V_0$ within the radius $r_0 = a/2$ and $0$ otherwise. $V_0$ is the square lattice potential depth. Note that the bold symbols (e.g. $\vec{r}$) represent 2D vectors. We define the characteristic energy scale as $E_0 = \frac{\hbar^2}{2m}|\frac{2\pi}{a}|^2 \sim 1$ meV for the $a$ = 4 $\mu$m device. Within the first BZ, the plane wave basis has a form of $|\vec{k}+\vec{G}_{mm}\rangle$ in terms of the 2D reciprocal lattice vectors $\vec{G}_{mn}$. Diagonalizing the Hamiltonian matrix $\langle\vec{k}+\vec{G}_{mm}|\hat{H}|\vec{k}+\vec{G}_{mm}\rangle$ with the operator $\hat{H}= - \frac{\hbar^2\hat{k}^2}{2m}+\hat{V}$ in the momentum space, the resulting bandstructures along $\Gamma, X, M$ are plotted in Fig.~\ref{fig:8}\textbf{b}. Figure~\ref{fig:8}\textbf{c} draws the first four BZs with locations of $\Gamma, X, M$ and the experimentally observed BZs are shown in Fig.~\ref{fig:8}\textbf{d} above threshold values. Because of the non-zero periodic potential, the degenerate eigenstates are lifted at the high symmetry points  and classified according to the rotational group symmetry analogous to the atomic orbital denotations. For example, at $\Gamma$ point, the lowest ground state exhibits the non-degenerate 1$s$-wave symmetry, whereas the next quartet states exhibit $3d_{x^2-y^2}, 2p_x, 2p_y,$ and $2s$-wave symmetries from the top to the bottom respectively. Similarly, at $M$ point, the degenerate quartet states are split into $3d_{xy}, 2p_x, 2p_y,$ and $1s$-wave symmetries.

Figure~\ref{fig:9}\textbf{a} shows the evolution of the LP momentum space population distributions taken from the far-field (FF) images. The regular square-patterned sharp interference peaks come from the diffracted laser signals from the 2D square lattices, which are useful to calibrate the momentum space. Since the device sits around  -3 meV detuned area, the donut shape,the bottleneck nature of the LP distribution in momentum space appears at $P/P_{th} \sim$ 0.29. As the injected polaritons are increased, the emission intensity grows at the four $M$ points, the corners of the first BZs. These peaks are further sharpened, and LP populations are further transferred to the $X$ points and ultimately relaxed to the inside of the first BZ. Furthermore, the states at $M$ points are stronger emissions under the metal films (red circles) in comparison to the lower energy states emit strongly through the holes (blue squares) from the near-field spectroscopy measurements above the threshold (Fig.~\ref{fig:9}\textbf{b}). Similar to the $p$-wave symmetry in the previous 1D section, we can conclude that the $M$ eigenstates are also anti-phased. And the coherence among $M$ points produces the interference patterns by the self-folding in the Michelson interferometer shown in Fig.~\ref{fig:9}\textbf{c}.

\begin{figure}[t]
\centering
\includegraphics*[width=.7\textwidth]{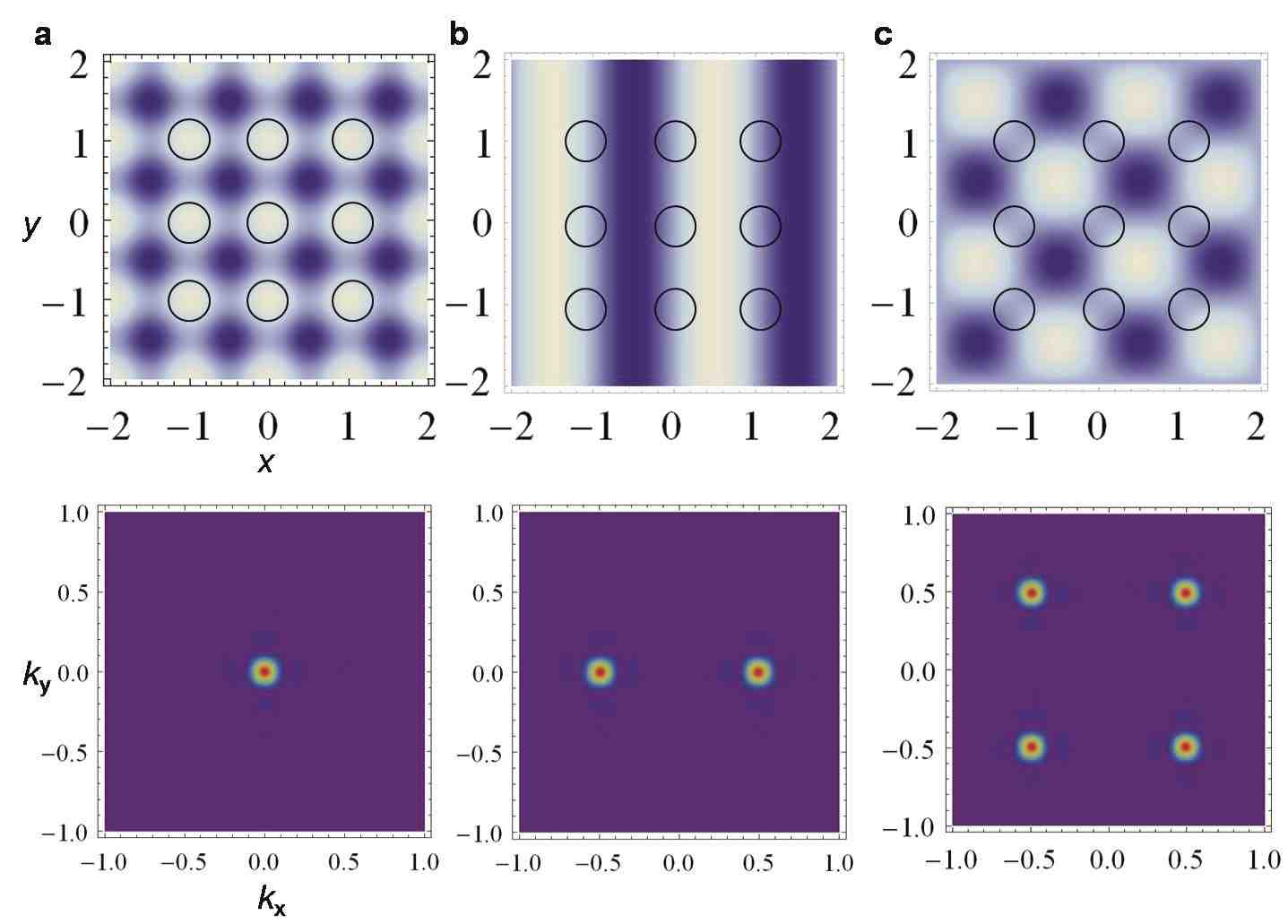}
\caption[]{Computed near-field lower polariton wavefunction (upper panel) and far-field intensities (lower panel) in the weak square lattice potential. 1$s$- (a), 2$p_x$- (b) and 3$d_{xy}$- (c) orbital condensates are presented. The black dotted circles are where the circular apertures are located in the metal film pattern. Whereas the 1s wavefunction is in-phased (a), both 2$p_x$ and 3$d_{xy}$ orbitals are anti-phased, inducing the strong interference peaks at $X$ and $M$ points in the far-field images. Permission of figure from Ref.~\cite{Kim2011} is obtained from Nature Physics}
\label{fig:10}       
\end{figure}

Figure~\ref{fig:10} summarizes the theoretical near-field (NF) and the FF distributions of $1s$, 2$p_x$ and $3d_{xy}$ states in the 3-by-3 square lattice obtained by the previous single-particle calculation. Since the potential depth is relatively weak, the LP wavefunctions are rather extended in the 2D; however, it is still clear that both 2$p_x$ and 3$d_{xy}$ states are anti-phased in the NF space unlike the 1$s$ state. Due to the orthogonal square lattice geometry, the interference peaks at $M$ points emerge uniquely from the anti-phased 3$ d_{xy}$ states (Fig.~\ref{fig:10}\textbf{c}). Experimentally, we are able to access both NF and FF LP population information (Fig.~\ref{fig:9}) to nicely confirm the theoretical conjecture.

This is the first time to observe that the dynamic condensation takes place with the anti-phased $d_{xy}$-wave symmetry. Again, the advantages to create high-orbital condensate states in the exciton-polariton systems come from the competition between the leaked photons through the cavity mirrors and the thermal relaxation processes to the ground states. In addition, the compact photoluminescence setup using the Fourier optics allows to examine both the NF and FF images and spectroscopies. The high-orbital condensate state is the stepping stone to realize solid-state quantum simulators, which would explore macroscopic quantum order in designated solid-state systems with high-orbital electronic states.



\section{Outlook}

Solid-state microcavity exciton-polariton condensates are regarded as a promising physical system to explore the fundamental physical phenomena revealing quantum bose nature at much elevated temperatures and in non-equilibrium situations. There are active research activities to engineer the trap potential for condensates in a controlled and systematic manner.  Among numerous trapping schemes, our group has been using a simple metal-film deposition method to enjoy flexibility and scalability in multi-dimensions. We have observed high-orbital condensates in meta-stable states formed by given lattice bandstructures and short polariton lifetimes in the cavity. The capability to manipulate condensate orbital states selectively is essential to emulate real material systems. For example, the 2D CuO plane of the high-temperature superconductors mixes with the $d-$ and $p-$orbitals. Thus, we are equipped to arrange multi-orbital condensates in 2D. In addition, upon the successful demonstration in the 1D and 2D square lattices, it is straightforward to extend to other 2D lattices, for example, triangular, honeycomb and kagome geometries. The polariton-lattice system has shown a great potential to investigate exciting physical questions even including spin degrees of freedom, from which we would envision to grasp important clues towards the understanding of strongly correlated condensed matter systems.

\printindex

\begin{thebibliography}{99.}

\bibitem{Jackson} J. D. Jackson, \textit{Classical Electrodynamics}, 2nd edn. (John Wiley \& Sons, Hoboken, N. J., 1975)

\bibitem{Haroche1989} S. Haroche and D. Kleppner, Physics Today \textbf{42}, 24 (1989)
\bibitem{Vahala2003} K. J. Vahala, Nature \textbf{424}, 839 (2003)
\bibitem{Weisbuch1992} C. Weisbuch et al., Phys. Rev. Lett. \textbf{69}, 3314 (1992)
\bibitem{Kavokin} A. Kavokin, \textit{Cavity Polaritons}, Volume 32 (Academic Press, 2003)
\bibitem{Yamamoto} Y. Yamamoto, F. Tassone, and H. Cao, \textit{Semiconductor Cavity Quantum Electrodynamics}, 1st edn. (Springer, 2000)
\bibitem{Deng2010} H. Deng, H. Haug, and Y. Yamamoto, Rev. Mod. Phys. \textbf{82}, 1489 (2010)
\bibitem{Imamoglu1996} A. Imamoglu et al., Phys. Rev. A \textbf{53}, 4259 (1996)
\bibitem{Deng2002} H. Deng et al., Science \textbf{298}, 199 (2002)
\bibitem{Deng2003} H. Deng et al., Proc. Nat. Acad. Sci. \textbf{100}, 15318 (2003)
\bibitem{Kasprzak2006} J. Kapsrzak et al., Nature \textbf{443}, 409 (2006)
\bibitem{Deng2006} H. Deng et al., Phys. Rev. Lett. \textbf{97}, 146402 (2006)
\bibitem{Deng2007} H. Deng et al., Phys. Rev. Lett. \textbf{99}, 126403 (2007)
\bibitem{Balili2007} R. B. Balili et al., Science \textbf{316}, 1007 (2007)
\bibitem{Lai2007} C.-W. Lai et al., Nature \textbf{450}, 529 (2007) 
\bibitem{Christopoulos2007} S. Christopoulos et al., Phys. Rev. Lett. \textbf{98}, 126405 (2007)
\bibitem{Christmann2008} G. Christmann et al., Appl. Phys. Lett. \textbf{93}, 051102 (2008)
\bibitem{Cohen2010} S. Kena Cohen and S. R. Forrest, Nature Photon. \textbf{4}, 371 (2010)
\bibitem{Mermin1966} N. D. Mermin and H. Wagner, Phys. Rev. Lett. \textbf{17}, 3314 (1133)
\bibitem{Hohenberg1967} P. C. Hohenberg Phys. Rev. \textbf{158}, 383 (1967)
\bibitem{Kosterlitz1973} J. M. Kosterlitz and D. J. Thouless, J. Phys. C: Solid State Phys. \textbf{6},1181 (1973)
\bibitem{Kosterlitz1974} J. M. Kosterlitz, J. Phys. C: Solid State Phys. \textbf{7}, 1046 (1974)
\bibitem{Berezinskii1971} V. L. Berezinskii, Zh. Eksp. Teor, Fiz. \textbf{61}, 1144 (1971) (Engl. Trans. Sov. Phys.-JETP \textbf{34}, 610 (1972))
\bibitem{Lagoudakis2008} K. G. Lagoudakis et al., Nat. Phys. \textbf{4}, 706 (2008)
\bibitem{Lagoudakis2009} K. G. Lagoudaki et al., Science \textbf{326}, 974 (2009)
\bibitem{Sanvitto2009} D. Sanvitto et al., Phys. Rev. B \textbf{80}, 045301 (2009) 
\bibitem{Roumpos2010} G. Roumpos et al., Phys. Rev. Lett. \textbf{104}, 126403 (2010) 
\bibitem{Bloch1998} J. Bloch et al., Physica E \textbf{2}, 915 (1998) 
\bibitem{Gutbrod1998} T. Gutbrod et al., Phys. Rev. B \textbf{57}, 9950 (1998)
\bibitem{Obert2004} M. Obert et al. Appl. Phys. Lett. \textbf{84}, 1435 (2004)
\bibitem{Bajoni2008} D. Bajoni et al., Phys. Rev. Lett. \textbf{100}, 047401 (2008) 
\bibitem{Ferrier2011} L. Ferrier et al., Phys. Rev. Lett. \textbf{106}, 126401 (2011)
\bibitem{Wertz2010} E. Wertz et al., Nat. Phys. \textbf{6},860 (2010) 
\bibitem{Galbiati2011} M. Galbiati, arXiv:1110.0359 (2011)
\bibitem{ElDaif2006} O. El Da\"if et al., Appl. Phys. Lett. \textbf{88}, 061105 (2006) 
\bibitem{Kaitouni2006} R. Idrissi Kaitouni et al., Phys. Rev. B \textbf{74}, 155311 (2006)
\bibitem{Nardin2010} G. Nardin et al., Phys. Rev. B \textbf{82}, 045304 (2007)
\bibitem{Fisher1995} T. A. Fisher et al., Phys. Rev. B \textbf{51}, 2600 (1995)
\bibitem{Miller1985} D. A. B. Miller et al., Phys. Rev. B \textbf{32}, 1043  (1985)
\bibitem{Balili2006} R. B. Balili et al., Appl. Phys. Lett. \textbf{88}, 031110 (2006) 
\bibitem{Balili2009} R. Balili et al., Phys. Rev. B \textbf{79}, 075319 (2009) 
\bibitem{Nelsen2009} B. Nelsen et al., J. Appl. Phys. \textbf{105}, 122414 (2009) 
\bibitem{Kim2008} N. Y. Kim et al., Phys. Stat. Solidi B \textbf{245}, 1076 (2008)
\bibitem{Yeh}P. Yeh, \textit{Optical waves in layered media}, (John Wiley \& Sons, Hoboken, N. J., 2005)
\bibitem{Bogoliubov1947} N. N. Bogoliubov, J. Phys. \textbf{11}, 23 (1947)
\bibitem{Pitaevskii} L. P. Pitaevskii and S. Stringari, \textit{Bose-Einstein Condensation}, (Clarendon Press, 2004)

\bibitem{Kim2011} N. Y. Kim et al., Nat. Phys. \textbf{7}, 681 (2011) 
\bibitem{Utsunomiya2008} S. Utsunomiya et al., Nat. Phys. \textbf{4},700 (2008) 
\bibitem{Lai2007} C.-W. Lai et al., Nature \textbf{450}, 529 (2007) 



\bibitem{Ashcroft} N. W. Ashcroft and N. D. Mermin, \textit{Solid State Physics}, (Brooks Cole, New York, 1989)

\bibitem{Cerda2010} E. A. Cerda-M$\acute{e}$ndez et al., Phys. Rev. Lett. \textbf{105}, 116402 (2010) 
\bibitem{deLima2006} M. M. de Lima, Jr. et al., Phys. Rev. Lett. \textbf{97},045501 (2006) 
\bibitem{Bloch1998} J. Bloch et al., Physica E \textbf{2}, 915 (1998) 





\bibitem{Liew2011} T. C. H. Liew, I. A. Shelykh, and G. Malpuech, Physica E \textbf{43}, 1543 (2011)
\bibitem{Deveaud2008} B. Deveaud-Pl$\acute{e}$dran, Nature \textbf{453}, 297 (2008)






























\end{thebibliography}
\end{document}